\documentclass[fleqn,twoside]{article}
\usepackage{espcrc2}
\usepackage{epsfig}

\def\ee{\mbox{e}^+\mbox{e}^-}
\def\ttbar{\mbox{t}\overline{\mbox t}}
\def\bbbar{\mbox{b}\overline{\mbox b}}
\def\ccbar{\mbox{c}\overline{\mbox c}}
\def\ffbar{\mbox{f}\overline{\mbox f}}
\def\QQbar{\mbox{Q}\overline{\mbox Q}}

\usepackage{graphicx}
\usepackage[figuresright]{rotating}

\newcommand{\AmS}{{\protect\the\textfont2
  A\kern-.1667em\lower.5ex\hbox{M}\kern-.125emS}}

\title{Physics at a Photon Collider}

\author{Stefan S\"oldner-Rembold\address[MCSD]{
        Fermi National Accelerator Laboratory,
        P.O. Box 500, Batavia, USA}
\thanks{Talk given at SUSY2002 and ICHEP02}
}%
       
\begin{document}

\begin{abstract}
A Photon Collider will provide unique opportunities to
study the SM Higgs boson and to determine its properties.
MSSM Higgs bosons can be discovered at the Photon Collider
for scenarios where they might escape detection at the LHC.
As an example for the many other physics topics which
can be studied at a Photon Collider, recent results on
Non-Commutative Field Theories are also discussed.
\vspace{1pc}
\end{abstract}

\maketitle

\section{Introduction}
The Photon Collider option of a Linear Collider (LC)
is based on laser photon back-scattering on 
high energy electrons. The maximum photon energy is 205~GeV for 
a laser with $\lambda=1.06 \mu\mbox{m}$ and an electron beam
energy of 250~GeV. A high degree of polarisation with 
opposite helicities of the electron and the laser photon
is crucial for obtaining a peaked spectrum of high energy polarised photons 
close to the maximum energy. In this case the high energy part of the
$\gamma\gamma$ spectrum is dominated by the spin-0
configuration which is important to enhance the signal and suppress
the background for Higgs production.
Alternatively, e$\gamma$ interactions are also possible. 
The technical aspects of the photon collider are discussed 
in~\cite{bib-jeff,bib-tdr}.

\section{Higgs Production}
Neutral Higgs bosons are produced in the scattering of two photons
as a $s$-channel resonance
through a loop. In this loop all charged particles contribute which obtain
their mass from electroweak symmetry breaking. The two-photon partial width
of the Higgs boson is therefore sensitive to physics beyond the SM.

For Higgs bosons decaying predominantly into $\bbbar$, the main
source of background are $\gamma\gamma\to\QQbar$ processes (Q=c,b).
The spin-0 component of these processes is suppressed in Leading Order (LO)
by a factor $m^2_Q/s$. Since this suppression is only valid in LO,
a realistic background simulation should be based on a next-to-leading (NLO)
calculation. 

\begin{figure}[htbp]
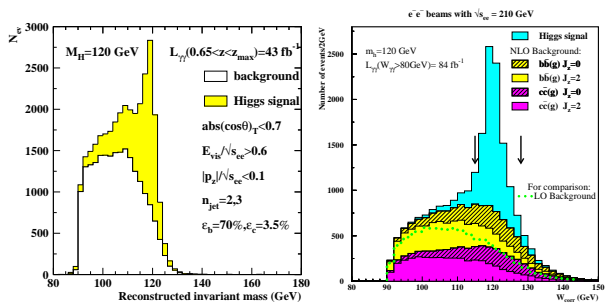

   \begin{center}
 \label{fig1}
\hspace{-8mm}
      \mbox{
          \epsfxsize=4.0cm
          \epsffile{fig5_120.epsi}
          \epsfxsize=3.8cm
          \epsffile{wcorr.epsi}
           }
   \end{center}
\vspace{-6mm}
\caption{Reconstructed invariant mass for a Higgs boson mass
of $120$~GeV with the full NLO background simulation.
The detector response is simulated with SIMDET. 
Left plot from~\protect\cite{bib-ssr} and right plot 
from~\protect\cite{bib-nzkbb}.
}
\label{fig-wcorr}
\end{figure}

Several analyses of $\mbox{H}\to\bbbar$ decays for a SM Higgs boson
in the mass range from 120 GeV to 160 GeV have been 
performed~\cite{bib-ssr,bib-nzkbb}.
All analyses exploit the kinematic differences between the
$s$-channel signal and the $t$-channel background by cutting
on visible energy and angular distributions.
The results are shown in Fig.~\ref{fig-wcorr}.

Before b-tagging the ratio of background from $\gamma\gamma\to\ccbar$ events
to $\gamma\gamma\to\bbbar$ events is approximately 16 due to the quark charges.
Excellent b-tagging is therefore required to suppress the
charm background. 
The need to minimize the radius of the beam pipe is one of the
main challenges for a Photon Collider, since the beam pipe 
has to accomodate the optical system for producing back-scattered 
photons.

The analyses presented here have assumed that $\bbbar$ events
are tagged with $70\%$ efficiency and $\ccbar$ events with $3.5\%$
efficiency. For a luminosity corresponding to roughly one year
of running, a statistical uncertainty of about $2\%$ for
a Higgs mass of 120-140~GeV for the two photon width 
measurement can be achieved. 
The uncertainty increases to about $10\%$ for a Higgs mass of 160~GeV. 

\begin{figure}[htbp]
   \begin{center}
      \mbox{
          \epsfxsize=7.0cm
          \epsffile{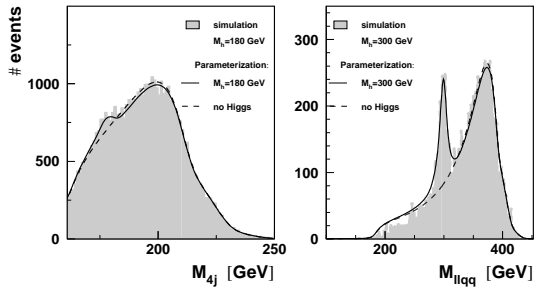}
           }
   \end{center}
\vspace{-6mm}
\caption{Reconstructed invariant mass for 
$\gamma\gamma\to {WW}$ events with a SM Higgs mass of 180~GeV (left) 
and for $\gamma\gamma\to \mbox{ZZ}$ events with a SM Higgs mass of 
300~GeV (right)~\protect\cite{bib-nzkww}.}
\label{fig-ww}
\end{figure}

At Higgs boson masses above 160~GeV decays into WW and ZZ 
pairs become important.
In this case the interference between signal and $\gamma\gamma
\to\mbox{WW}$ background needs to be taken into account. 
The interference gives access to an additional observable, 
the phase $\phi_{\gamma\gamma}$ of the $\gamma\gamma\to\mbox{H}$ amplitude.
The combined precision of phase and partial width determination
gives sufficient precision to distinguish the SM from 
SM-like 2HDM (II) scenarios~\cite{bib-nzkww}. 
The results of a detector simulation
are shown in Fig.~\ref{fig-ww}.

\begin{figure}[htbp]
   \begin{center}
      \mbox{
          \epsfxsize=6.5cm
          \epsffile{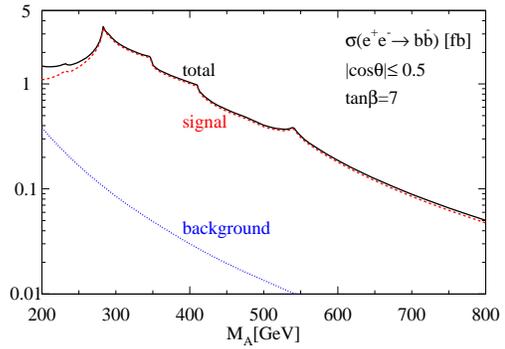}
           }
   \end{center}
\vspace{-6mm}
\caption{Cross-section for the process $\gamma\gamma\to\mbox{A}\to\bbbar$
and for the background $\gamma\gamma\to\bbbar$. A mass window of $\pm 3$~GeV
has been applied, $100\%$ polarisation is assumed, and only the two-jet
configuration is considered~\protect\cite{bib-mssm}.
   }
\label{fig-mm}
\end{figure}

The neutral MSSM Higgs Boson H,A for masses above 200~GeV
and for moderate $\tan\beta\approx 7$ might escape detection
at the LHC. In this parameter region, where decays into
$\bbbar$ are the dominant SM decays up to Higgs masses around 550 GeV,
the Photon Collider can discover the neutral MSSM Higgs 
Bosons~\cite{bib-mssm}.
In contrast to the $\ee$ option of the LC, the Photon Collider can
produce these Higgs Boson with masses up to about $80\%$
of $\sqrt{s}_{\rm ee}$. Cross-sections for signal and
background are shown in Fig.~\ref{fig-mm}.

Many other Higgs scenarios have been studied for the Photon Collider
option, adding to the discovery potential: 
For example, the measurement of CP properties of the Higgs bosons
A,H in $\ttbar$ decays~\cite{bib-tt} or the production of
charged Higgs bosons in the process 
$\gamma\gamma\to\mbox{H}^{+}\mbox{H}^{-}$~\cite{bib-asner}.

\section{Non-Commutative Field Theories}
\begin{figure}[htbp]
   \begin{center}
      \mbox{
          \epsfxsize=3.5cm
          \epsffile{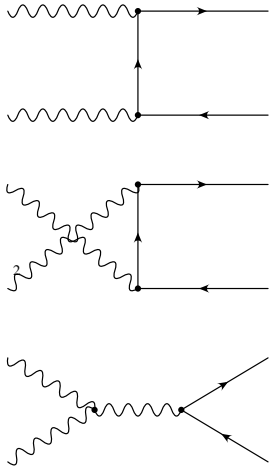}
           }
   \end{center}
\vspace{-6mm}
\caption{Diagrams contributing to fermion pair production
($\gamma\gamma\to\ffbar$) in NCQFT.}
\label{fig-qed}
\end{figure}

One of many other interesting topics which can be studied
at a $\gamma\gamma$ or at an e$\gamma$ collider are 
non-commutative quantum field theories (NCQFT)
with non-commuting (NC) space-time operators~\cite{bib-god2}.
The additional cubic coupling ($\gamma\gamma\gamma$) contributing
to the process $\gamma\gamma\to\ffbar$ is shown in Fig.~\ref{fig-qed}.

\begin{figure}[htbp]
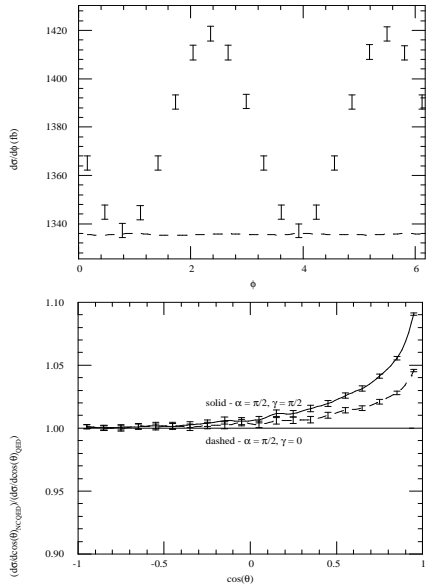

   \begin{center}
          \mbox{
          \epsfxsize=3.8cm
          \begin{turn}{-90}
          \epsffile{fig10b.epsi}
          \epsfxsize=3.8cm
          \epsffile{NCQFT.epsi}
          \end{turn}
           }
   \end{center}
\vspace{-6mm}
\caption{$\mbox{e}\gamma\to\mbox{e}\gamma$ scattering at $\sqrt{s}_{ee}
=500$~GeV
a) Differential cross-section ${\rm d}{\sigma}/{\rm d}\phi$ for
the SM (dashed line) and for $\Lambda_{\rm NC}=500$~GeV 
($\alpha=\gamma=\pi/2$). 
The expected statistical uncertainties are also shown.
b) Ratio of the differential cross-sections 
${\rm d}{\sigma}/{\rm d}\cos{\theta}$ for NCQFT and for the SM.
}
\label{fig-don}
\end{figure}
The parameter $\Lambda_{\rm NC}$ characterises the threshold
where NC effects become relevant. The current limit from
$\ee$ scattering is $\Lambda_{\rm NC}>142$~GeV at $95\%$
confidence level~\cite{bib-pn500}.

A theoretical analysis has been performed of the processes
$\gamma\gamma\to\ffbar$ and $\mbox{e}\gamma\to\mbox{e}\gamma$
for $L_{\rm ee}=500$~fb$^{-1}$. A transverse momentum greater
than 10~GeV and a polar angle in the range $10^{\circ}<\theta<170^{\circ}$
has been required for the final state particles.

In Fig.~\ref{fig-don} the NCQFT effects on the angular distributions
of the final state photons in $\mbox{e}\gamma\to\mbox{e}\gamma$ scattering
are shown. The parameters $\alpha$ and $\gamma$ are related to
the Maxwell field tensor~\cite{bib-god2}.
Significant deviations from the SM can be observed.

\section{Conclusion}
Studies of various Higgs scenarios
show that a Photon Collider has a unique potential
for Higgs boson physics over a wide
range of masses and model parameters.
Excellent b-tagging and good energy
resolution are very important for the Photon Collider to
suppress background.
The e$\gamma$ option of the Photon Collider is complementary to
the $\ee$ Linear Collider in discovering effects from
Non-Commutative Field Theories. Many other topics which
can be studied at a Photon Collider (e.g. SUSY, Leptoquarks, QCD)
had to be omitted in this short summary.


\vspace{-3mm}
\begin{thebibliography}{9}
\bibitem{bib-jeff}  J. Gronberg, these proceedings.
\bibitem{bib-tdr}   ECFA/DESY Photon Collider Working Group, 
                    hep-ex/0108012 (ABS 770).
\bibitem{bib-ssr}   G. Jikia, S. S\"oldner-Rembold,
                    Nucl. Inst. and Meth. A472 (2001) 133; 
                    Nucl. Phys. Proc. Suppl. 82 (2000) 373 (ABS 812).
\bibitem{bib-nzkbb} P. Nie\.zurawski, A.F. \.Zarnecki and M. Krawczyk, 
                    hep-ph/0208234 (ABS 665).
\bibitem{bib-nzkww} P. Nie\.zurawski, A.F. \.Zarnecki and M. Krawczyk, 
                    hep-ph/0207294 (ABS 155).
\bibitem{bib-mssm}  M. M\"uhlleitner et al., Phys. Lett. B508 (2001) 311;
                    D. Asner et al., hep-ph/0110320.
\bibitem{bib-tt}    E. Asakawa, hep-ph/0101234.
\bibitem{bib-asner} D. Asner et al., hep-ph/0208219.
\bibitem{bib-god2}  S. Godfrey, M. Doncheski, hep-ph/0111147;
                    Phys. Rev. D65 (2002) 015005 (ABS 625).
\bibitem{bib-pn500} OPAL Physics Note PN500 (ABS 889).
\end{thebibliography}
\end{document}